\documentclass[12pt]{article}
\usepackage{setspace}
\usepackage{epic,eepic,epsfig,changebar,amsmath,sint_new}
\usepackage[american]{babel}
\usepackage{fullpage}

\def\rmO{{\rm O}}

\def\proof{\noindent{\sl Proof:}\kern0.6em}

\def\dual{\mathstrut^*\kern-0.1em}

\def\lvec#1{\setbox0=\hbox{$#1$}
    \setbox1=\hbox{$\scriptstyle\leftarrow$}
    #1\kern-\wd0\smash{
    \raise\ht0\hbox{$\raise1pt\hbox{$\scriptstyle\leftarrow$}$}}
    \kern-\wd1\kern\wd0}
\def\rvec#1{\setbox0=\hbox{$#1$}
    \setbox1=\hbox{$\scriptstyle\rightarrow$}
    #1\kern-\wd0\smash{
    \raise\ht0\hbox{$\raise1pt\hbox{$\scriptstyle\rightarrow$}$}}
    \kern-\wd1\kern\wd0}

\def\nabstar#1{\nabla\kern-0.5pt\smash{\raise 4.5pt\hbox{$\ast$}}
               \kern-4.5pt_{#1}}

\def\drvstar#1{\partial\kern-0.5pt\smash{\raise 4.5pt\hbox{$\ast$}}
               \kern-5.0pt_{#1}}

\def\MeV{{\rm MeV}}

\def\fm{{\rm fm}}

\def\rhoprime{\rho\kern1pt'}
\def\rhobar{\bar{\rho}}
\def\rhobarprime{\rhobar\kern1pt'}
\def\rhobartilde{\kern2pt\tilde{\kern-2pt\rhobar}}
\def\rhobartildeprime{\kern2pt\tilde{\kern-2pt\rhobar}\kern1pt'}

\def\zetabar{\bar{\zeta}}
\def\zetaprime{\zeta\kern1pt'}
\def\zetabarprime{\zetabar\kern1pt'}
\def\zetar{\zeta_{\raise-1pt\hbox{\sixrm R}}}
\def\zetabarr{\zetabar_{\raise-1pt\hbox{\sixrm R}}}

\def\phiimpr{\phi_{\kern0.5pt\hbox{\sixrm I}}}

\def\diracstar#1#2{
    \setbox0=\hbox{$\gamma$}\setbox1=\hbox{$\gamma_{#1}$}
    \gamma_{#1}\kern-\wd1\kern\wd0
    \smash{\raise4.5pt\hbox{$\scriptstyle#2$}}}

\def\ba{b_{\rm A}}

\def\ca{c_{\rm A}}

\def\csw{c_{\rm sw}}

\def\fa{f_{\rm A}}

\def\f1{f_1}

\def\opprime#1{\setbox0=\hbox{${\cal O}$}\setbox1=\hbox{${\cal O}_{\rm #1}$}
    {\cal O}_{\rm #1}\kern-\wd1\kern\wd0
    \smash{\raise4.5pt\hbox{\kern1pt$\scriptstyle\prime$}}\kern1pt}

\def\ophatprime#1{\setbox0=\hbox{$\widehat{\cal O}$}
    \setbox1=\hbox{$\widehat{\cal O}_{\rm #1}$}
    \widehat{\cal O}_{\rm #1}\kern-\wd1\kern\wd0
    \smash{\raise4.5pt\hbox{\kern1pt$\scriptstyle\prime$}}\kern1pt}

\def\bopprime#1{\setbox0=\hbox{${\cal O}$}\setbox1=\hbox{${\cal O}_{\rm #1}$}
    {\cal L}_{\rm #1}\kern-\wd1\kern\wd0
    \smash{\raise4.5pt\hbox{\kern1pt$\scriptstyle\prime$}}\kern1pt}

\def\blagprime#1{\setbox0=\hbox{${\cal B}$}\setbox1=\hbox{${\cal B}_{#1}$}
    {\cal B}_{#1}\kern-\wd1\kern\wd0
    \smash{\raise5.2pt\hbox{\kern1pt$\scriptstyle\prime$}}\kern1pt}

\def\za{Z_{\rm A}}

\def\msbar{{\rm \overline{MS\kern-0.05em}\kern0.05em}}
\def\MSbar{{\rm \overline{MS\kern-0.05em}\kern0.05em}}

\def\mDs{m_{\rm D_s}}

\def\Ds{\rm D_s}
\def\fDs{F_{\Ds}}

\begin{document}
\begin{titlepage}
\begin{flushright}
   HU-EP-03-09
\end{flushright}
\vskip 1 cm
\begin{center}
  {\Large\bf  A precise determination of the decay constant of the
  D$_{\text{s}}$-meson in quenched QCD}
\end{center}
\vskip 1 cm
\begin{figure}[h]
\begin{center}
\epsfig{figure=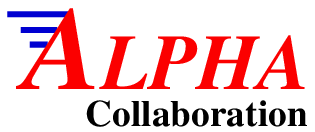} 
\end{center}
\end{figure}
\begin{center}
{\large Andreas J\"uttner\hskip .2ex$^{\scriptscriptstyle a}$ and
Juri Rolf\hskip .2ex$^{\scriptscriptstyle b}$}
\vskip 2.3ex
\begin{center}
Institut f\"ur Physik, Humboldt-Universit\"at zu Berlin\\
Invalidenstr. 110, D-10115 Berlin, Germany\\
Email: $^{\scriptstyle a}$\hskip .2ex juettner@physik.hu-berlin.de\\
$^{\scriptstyle b}$\hskip .2ex rolf@physik.hu-berlin.de\\
\end{center}

\vskip 1.5cm
{\bf Abstract}
\vskip 0.7ex
\end{center}
We present a lattice determination of the leptonic decay constant
$\fDs$ of the D$_{\text{s}}$-meson using its mass and $F_{\rm K}$ as
experimental input.  Setting the scale with $F_{\rm K}=160\,\MeV$, our
final result is $\fDs = 252(9)\,\MeV$. The error contains all
uncertainties apart from the quenched approximation. Setting the scale
with the nucleon mass instead leads to an decrease of $20(1)\,\MeV$
of the leptonic decay constant.
   \vspace{.5cm}\\


\vfill \eject
\end{titlepage}

\section{Introduction}

The $\Ds^+$-meson consists of a c and a $\bar{\rm s}$ quark. It is
stable in QCD and decays by
an emission of a W$^+$-boson into a lepton and a neutrino. The amplitude
of this decay is characterized by the decay constant $\fDs$ which is
defined by the QCD matrix element
\begin{equation}
  \langle 0\vert A_{\mu}(0)\vert \Ds(p)\rangle = i p_{\mu} \fDs
\end{equation}
of the axial current $A_{\mu} = \bar{\rm s}\gamma_{\mu}\gamma_5{\rm c}$.  

Given the CKM-matrix element $V_{\rm cs}$, $\fDs$ can be measured
experimentally by studying the branching ratio  
$ \text{BR}(\Ds\rightarrow l\nu)$.
The current status of the experimental determination of $\fDs$ has
been summarized in~\cite{expsummary}.
Currently, the most precise data come from
the ALEPH experiment~\cite{aleph}, $\fDs =
(285\pm 19\pm 40)\,\MeV$, and from
CLEO~\cite{cleo}, $\fDs = (280\pm 17\pm 25\pm 34)\,\MeV$.
In the next couple of years
CLEO aims at reducing their error to two percent precision~\cite{cleointernet}.

The status of lattice computations for
$\fDs$~\cite{Bernard:2000ht,Becirevic:2000kq,Maynard:2001zd} was
reviewed in~\cite{Draper:1998ms,Bernard:2000ki} with a world average
of $\fDs = 255\pm 30\,\MeV$. With QCD sum rules one currently gets
$\fDs = 235\pm 24\,\MeV$~\cite{Narison:2001pu,Penin:2001ux}.

The goal of this work is a computation on the lattice of the weak
decay constant $\fDs$ to 3 percent accuracy up to the quenched
approximation. All the systematic and statistical errors will be
analyzed and we will also estimate the size of the quenched scale
ambiguity.  A precise quenched calculation of $\fDs$ together with
precise experimental data supplies us with a test of quenched lattice
QCD. This is of importance for the determination of the B-meson decay
constant $f_{\rm B}$ from lattice QCD which is an essential input in
the measurement of the third generation CKM-matrix elements.

\section{Strategy}

The decay constant $\fDs$ is defined in terms of a QCD matrix element.
To evaluate this matrix element we have to eliminate the bare
parameters of the QCD Lagrangian, that is the bare gauge coupling
$g_0$ and the bare masses of the relevant quarks, in favour of
physical observables.  One possible hadronic scheme, which has been
summarized in~\cite{charm} and developed
in~\cite{Capitani:1999mq,Garden:2000fg,charm} is to use the decay
constant $F_{\rm K}=160\,\MeV$ to set the scale\footnote{i.e.~to
  compute the lattice spacing $a$ in physical units as a function of
  the bare coupling} and eliminate the bare masses of the strange and
charm quarks in favour of the masses of the K and the $\Ds$-meson.
Neglecting isospin breaking, the quark mass ratio $M_s/\hat{M} =
24.4\pm 1.5$ with $\hat{M} = \frac{1}{2}(M_u+M_d)$ is taken from
chiral perturbation theory~\cite{Leutwyler:1996qg}.

The results in~\cite{Capitani:1999mq,Garden:2000fg,charm} are given in
terms of the Sommer scale $r_0$ which is derived from the force
between static colour sources. $r_0/a$ has been computed as a function
of the bare coupling $g_0$ to a high precision for a wide range of
cutoff values~\cite{Guagnelli:1998ud,Edwards:1997xf,Necco:2001xg}.  It
is only affected by lattice artifacts of ${\rm O}(a^2)$ in the
quenched theory~\cite{Necco:2001xg}. The relation to other hadronic
scales is known~\cite{Garden:2000fg}, $r_0\ F_{\rm K} = 0.415(9)$, and
$r_0\ m_{N}\approx 2.6$.  This illustrates the inconsistency of the
quenched approximation since the ratio $F_{\rm K}/m_N$ deviates by
approximately 10$\%$ from its experimental value. This is the typical
size of the quenched scale ambiguity. Setting the scale with $F_{\rm
  K}=160\,\MeV$ corresponds to setting $r_0=0.5\,\fm$ while $m_N =
938\,\MeV$ corresponds to $r_0=0.55\,\fm$.  Below we will estimate the
size of the quenched scale ambiguity for the weak decay constant of
the $\Ds$-meson.

This work aims at a scaling study of $\fDs$ on the lattice. Thus we
keep the physical conditions constant and only vary the lattice
spacing. 
We perform numerical simulations
in ${\rm O}(a)$ improved lattice QCD using Schr\"odinger functional boundary
conditions~\cite{Luscher:1992an,Sint:1993un} on a $L^3\times T$ space
time cylinder.  The 
practical advantages of this approach are discussed
in~\cite{Guagnelli:1999zf}. For unexplained notation
we refer to~\cite{Luscher:1996sc}.

In particular we define the meson sources 
\begin{equation}
{\cal O} = a^6\sum_{{\bf y}, {\bf z}}\overline{\zeta}_j({\bf
  y})\gamma_5\zeta_i({\bf z}),\quad 
{\cal O}' = a^6\sum_{{\bf y}, {\bf z}}\overline{\zeta}'_j({\bf
  y})\gamma_5\zeta'_i({\bf z}), 
\end{equation}
with flavour indices $i\neq j$ at the $x_0=0$ and the $x_0=T$ boundary
time slices, respectively. The correlation functions 
\begin{eqnarray}
  \label{eq:fA}
  f_{\rm A}(x_0) = -\frac{1}{2}\langle {\cal O} A_0(x)\rangle,\ \
  f_{\rm P}(x_0) = -\frac{1}{2}\langle {\cal O} P(x)\rangle,\ \
  f_1 = - \frac{1}{3L^6} \langle {\cal O}' {\cal O}\rangle
\end{eqnarray}
are then used to compute the decay constant $\fDs$ at finite lattice
spacing. Here $A_{\mu}(x) =
\overline{\psi}_i(x)\gamma_{\mu}\gamma_5\psi_j(x)$ is the axial
current and $P(x)= \overline{\psi}_i(x)\gamma_5\psi_j(x)$ is the
pseudoscalar density. The leading $\rmO(a)$ cutoff effects in $f_{\rm
  A}$ are canceled by using a clover improved
action~\cite{Sheikholeslami:1985ij} with coefficient $\csw$ and by
improving the axial current,
\begin{equation}
   A_{\mu}^I(x) = A_{\mu}(x) + \ca a \tilde{\partial}_{\mu} P(x).
\end{equation}
Both coefficients $\csw$ and $\ca$ are known non perturbatively in the
quenched approximation for bare couplings $g_0^2 = 6/\beta \leq
1$~\cite{Luscher:1996ug}. The improved correlation function $f_{\rm
  A}^I$ is then defined in analogy to $f_{\rm A}$.

On the lattice the axial current 
receives a multiplicative renormalization by the finite factor $\za$.
The renormalized axial current is thus given by
\begin{equation}
  (A_{\rm R})_{\mu} = \za \left(1 + \ba (am_{{\rm q},i} + am_{{\rm
  q},j})/2 \right)\,A_{\mu}^I + \rmO(a^2), 
\end{equation}
where the $\rmO(a)$ artifacts that are proportional to the bare
subtracted quark masses $am_{\rm q}$ are canceled by the term
proportional to $\ba$.

The correlation functions $f_{\rm A}^I$ and $f_{\rm P}$ decay
exponentially proportional to $e^{-x_0m_{\Ds}}$ when the excitations
due to higher states are small enough. Thus the meson mass $m_{\Ds}$,
which has the experimental value $r_0m_{\Ds} =
4.988$~\cite{PDG_review}, can be obtained from the plateau average of
the effective pseudoscalar mass
\begin{equation}
\label{eq:meff}
  am_{\text{PS}}(x_0+a/2) = \log\frac{f_{\rm A}^I(x_0)}{f_{\rm
  A}^I(x_0+a)}. 
\end{equation}
The weak decay constant of the $\Ds$-meson at finite lattice spacing
can then be obtained through \footnote{In~\cite{Guagnelli:1999zf} the
  first minus sign is missing.}
\begin{eqnarray}
\label{eq:fds}
  \fDs &=& -2 \za (1+\ba (am_{{\rm q},i}+am_{{\rm q},j})/2 )\
  \frac{\fa^I}{\sqrt{f_1}} \
  (m_{\Ds}L^3)^{-1^/2} e^{(x_0-T/2) m_{\Ds}} \nonumber\\
 &&\times 
  \left\lbrace 1 - \eta_A^{\Ds} e^{-x_0\Delta} - \eta_A^0
    e^{-(T-x_0) m_{\text{G}}}\right\rbrace + \rmO(a^2).
\end{eqnarray}
Here the factor $(m_{\Ds}L^3)^{-1^/2}$ takes into account the
normalization of one particle states.  The contribution $f_1^{-1/2}$
cancels out the dependence on the meson sources.  Because of the
exponential decay of the correlation function $\fa^I$ the product
in~(\ref{eq:fds}) is expected to exhibit a plateau at intermediate
times when the contribution $\eta_A^{\Ds} e^{-x_0\Delta}$ of the first
excited state and the contribution $\eta_A^0 e^{-(T-x_0)
  m_{\text{G}}}$ from the $O^{++}$ glueball both are small.  A plateau
average can then be performed to increase the signal and is understood
in~(\ref{eq:fds}).  Further explanations for equation~(\ref{eq:fds})
and details can be found in~\cite{Guagnelli:1999zf}.

\section{Numerical results}

\subsection{Parameters}

Our choice of simulation parameters is shown in
table~\ref{tab:sizes}. 
\begin{table}[htbp]
  \begin{center}
    \begin{tabular}{|l|l|ll|lll|l|}\hline
      $\beta$ & $n_{\text{meas}}$ & $L/a$ & $L/r_0$ & $\kappa_{\text{critical}}$ & $\kappa_s$ & $\kappa_c$& $r_0 m_{\Ds}$ \\ \hline 
      6.0     & 380               & 16    &   2.98  & 0.135196 & 0.133929 &  0.119053   & 4.972(22)     \\
      6.1     & 301               & 24    &   3.79  & 0.135496 & 0.134439 &  0.122490   & 4.981(23)     \\
      6.2     & 251               & 24    &   3.26  & 0.135795 & 0.134832 &  0.124637   & 5.000(25)     \\
      6.45    & 289               & 32    &   3.06  & 0.135701 & 0.135124 &  0.128131   & 5.042(29)     \\ \hline
    \end{tabular}
    \caption{Statistics and parameters for our simulations and
      demonstration of constant physical conditions.}
    \label{tab:sizes}
  \end{center}
\end{table}
The bare couplings and the hopping parameters for the strange quark
are the same as used in~\cite{charm}, and the hopping parameters for
the charm quark have been found in that work. The non perturbatively
defined improvement coefficients $\csw$ and $\ca$ are taken
from~\cite{Luscher:1996ug}.  $\ba$ has been obtained non
perturbatively in~\cite{Bhattacharya:2001ks}. Since it is difficult to
compute (cmp.~\cite{Guagnelli:2000jw}) and since we have to
extrapolate their results slightly we have used one-loop perturbation
theory~\cite{Sint:1997jx} as well.  The conversion to physical units
is done with the Sommer scale $r_0$.
From the fit function of~\cite{Guagnelli:1998ud} we infer that our
lattice size $L/a=16$ at $\beta=6.0$ corresponds to $L/r_0=2.98$.  The
results of~\cite{Garden:2000fg} show that at this volume the finite
size effects may be neglected.  The lattice spacing $a$ is varied from
$0.1\,\fm$ to $0.05\,\fm$.

Furthermore we choose $T/L= 2$ in contrast to $T/L = 2.5$ used for the
computation of the charm quark mass in~\cite{charm} to reduce rounding
effects, which would be the dominant systematic effects on $\fDs$
otherwise.

The fourth column of table~\ref{tab:sizes} shows that the above choice
of simulation parameters corresponds to approximately constant
physical box sizes. Also the mass of the $\Ds$-meson takes its correct
value up to small fluctuations. We have checked that these lead to
negligible corrections for the decay constant $\fDs$.

\subsection{Computation of the $\Ds$-meson decay constant}

To compute the meson mass $m_{\Ds}$ and the decay constant $\fDs$ we
calculate the combinations~(\ref{eq:meff},\ref{eq:fds}) of correlation
functions. For all the parameter choices we find plateaus as functions
of $x_0$. Our task is to find the plateau region such that the effect
of excited states is negligible compared to our statistical error.

When dealing with heavy quark propagators on single precision machines
there is always the danger of roundoff problems which might be the
dominant uncontrolled systematic error. Since the computation of the
decay constant $\fDs$ involves propagators extending from the $x_0=0$
to the $x_0=T$ boundary in the correlation functions $f_1$ the
discussion of these issues becomes even more important as for the
computation of masses. From~\cite{charm} we estimate that for $T=2L$
instead of $T=2.5 L$ taken there the rounding errors for $\fDs$ should
be small enough.
To show this we computed our observables on ${\cal O}(10)$
configurations at $\beta=6.0$ and $6.45$ in single as well as in
double precision. This check reveals that the rounding errors for
$\fDs$ are of the order of one per mill.

From fits of $\fDs(x_0)$ using an a priori chosen plateau range we
extract the contributions $\eta_A^{\Ds} e^{-x_0 \Delta}$ and $\eta_A^0
e^{-(T-x_0) m_{\text{G}}}$ of the first excited state and the $0^{++}$
glueball excitation, respectively. The theory
in~\cite{Guagnelli:1999zf} predicts relations between the prefactors
and the corresponding prefactors for the pseudoscalar masses which are
roughly \footnote{In some cases the fit function is not determined
  precisely. Then we choose the fit function such that the
  contribution of the excited states is rather overestimated.  Note
  that we do not attempt to measure $\Delta$ or $\eta_A^{\Ds}$ but
  rather want to estimate our systematic effects.} fulfilled.

All the systematic effects that deteriorate the plateau are added.
With an a priori chosen upper bound $\delta$ for the maximal relative
systematic error in a point $\fDs(x_0)$ we define the time interval
for the plateau region such that for all times in this interval the
added systematic contribution is smaller than $\delta$.  The
systematic error for the plateau averaged observable is then even
smaller.  For the decay constant we took $\delta=0.5\%$.  The plateaus
defined in this way extend from $4r_0$ to $5r_0$ at the four $\beta$
values considered.  For the pseudoscalar mass we used $\delta=0.3\%$.

\subsection{Continuum limit}

\begin{figure}[htb]
  \begin{center}
\vspace{-5cm}
      \psfig{file=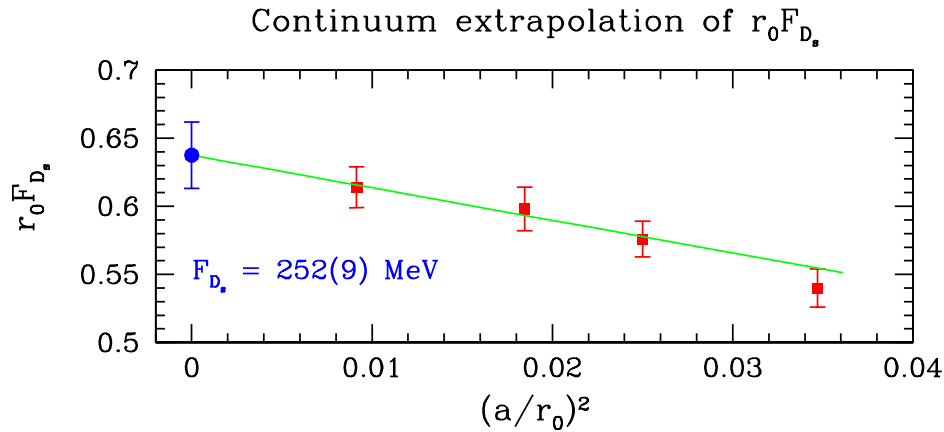,
      width=0.8\linewidth} 
    \caption{Continuum extrapolation of $\fDs$.}
    \label{fig:context}
  \end{center}
\end{figure}

We perform the plateau averages of~(\ref{eq:fds}) and obtain the
decay constant at finite lattice spacings as shown in table~\ref{tab:res}.
\begin{table}[htbp]
  \begin{center}
    \begin{tabular}{|l|l|l|l|}\hline
      $\beta$ & $a\fDs^{\text{bare}}\vert_{\ca=0}$ & $\partial
        a\fDs^{\text{bare}}/\partial \ca$ & $r_0 \fDs$ \\ \hline
        6.0 & 0.1144(25) & 0.2375(58) & 0.540(14) \\ 
        6.1 & 0.0979(17) & 0.1699(33) & 0.576(13) \\
        6.2 & 0.0863(20) & 0.1280(32) & 0.598(16) \\
        6.45& 0.0633(14) & 0.0664(17) & 0.614(15) \\ \hline
    \end{tabular}
  \end{center}
\caption{Simulation results for $\fDs$.}
\label{tab:res}
\end{table}
These data can be extrapolated to the continuum limit, leading to our
main result.  Since we employ ${\rm O}(a)$ improvement, the natural
ansatz for the continuum extrapolation is linear in $a^2$. This
extrapolation is shown in figure~\ref{fig:context}.
In the fit we neglect the point at $\beta=6.0$ which is farthest
away from the continuum limit. We quote 
\begin{equation}
  r_0 \fDs = 0.638(24)
\end{equation}
as our final result. Using $r_0=0.5\,\fm$ this corresponds to 
$\fDs = 252(9)\,\MeV$.
Using 1-loop perturbation theory for $\ba$ instead of the results
of~\cite{Bhattacharya:2001ks} leads to
\begin{equation}
  r_0 \fDs = 0.631(24),
\end{equation}
corresponding to $\fDs = 249(9)\,\MeV$.

\subsection{Quenched scale ambiguity}

To estimate the quenched scale ambiguity we consider $r_0\fDs$ as a
function $f(z)$ of $z=r_0\mDs$ around the physical value of $z$. This
is possible since in addition to the hopping parameters leading to the
physical strange and charm quark masses we have computed our
observables at five more hopping parameter values around the hopping
parameter of the charm quark for each bare
coupling considered. Here we will only use the two hopping parameters
closest to the charm value and present further results in a separate
publication. 

We expand $f(z) = f(z_0) + (z-z_0) f'(z_0) + \ldots$ in a Taylor
series around $z_0=4.988$. $f'(z_0)$ is estimated from a linear fit
of $r_0\fDs$ as a function of $r_0\mDs$. A 10$\%$ increase of $r_0$
corresponds to $z-z_0=0.5.$ Our estimate of the corresponding effect
on $r_0\fDs$ in the continuum limit is
\begin{equation} 
  0.5 f'(z_0) = 0.008(3).
\end{equation} 
Converting back to physical units (now using $r_0=0.55\,\fm$) we
estimate that $\fDs$ decreases by $20\,\MeV$ corresponding to eight
percent.

\section{Conclusions}

Our main result is $r_0\fDs=0.638(24)$. Using $r_0=0.5\,\fm$ this
corresponds to $\fDs=252(9)\,\MeV$. This is compatible with all the
experimental data currently available and with most theoretical
estimates as well. The precision of our result however matches the
goal of precision of future experiments, for example CLEO.

Under a scale shift of 10$\%$, which is typical for quenched spectrum
computations $\fDs$ decreases by eight percent. Assuming
that all the scale ambiguity comes from neglecting dynamical sea quark
effects this might also give an idea of the unquenching effect for
$\fDs$ which however has to be probed in an unquenched simulation.

Since our complete data set consists also of simulations at larger
quark masses we will be able to look at the functional dependence of
the weak decay constant around the charm quark mass. Together with a
new computation of the decay constant in the static
approximation we will then interpolate to the $B$-meson mass
and compute $F_{\rm B}$ with a better precision than is currently
available. The corresponding analysis has still to be done and further
simulations will be necessary so that these exciting results will be
presented in another publication.

\vskip 1ex This work is part of the ALPHA collaboration research
programme. It was supported by the European Community under the grant
HPRN-CT-2000-00145 Hadrons/Lattice QCD and by the Deutsche
Forschungsgemeinschaft in the SFB/TR 09 and the Graduiertenkolleg
GK271.  All the production runs were carried out on machines of the
APE1000 series at DESY-Zeuthen. The checks of the rounding errors in
double precision were done with a C-code based on the MILC
collaboration's public lattice gauge theory code, see~\cite{milc}.
This code ran on the IBM p690 system of the HLRN~\cite{hlrn} and on
the PC cluster at DESY-Zeuthen.  We thank the staff at the computer
centres for their help, and M.~Della Morte, S.~Sint, R.~Sommer, and
U.~Wolff for useful discussions.

\end{document}